\documentclass[aps,prx,twocolumn,floats,showpacs,superscriptaddress,nofootinbib]{revtex4-1}
\usepackage{longtable}
\usepackage{graphicx,epsfig}
\usepackage[normalem]{ulem}
\usepackage{times,bbm}
\usepackage{graphics,dcolumn,bm,float}
\usepackage{amssymb,amsmath,rotate,color,amsfonts}
\usepackage[title,titletoc,toc]{appendix}
\usepackage{mathtools}
\usepackage{booktabs}
\usepackage{tcolorbox}
\usepackage[pagebackref=false,colorlinks,linkcolor=magenta,citecolor=blue,urlcolor=magenta]{hyperref}
\usepackage[all]{hypcap}

\usepackage{wrapfig}
\usepackage{lipsum}
\usepackage{mwe}
\usepackage[mathlines]{lineno}
\usepackage[mathscr]{euscript}
\usepackage{hyperref}
\usepackage{breqn}
\usepackage{bbold}
\usepackage{pgfplots}
\usepackage{float}
\usepackage{tkz-euclide}
\usepackage{braket}
\usepackage{physics}
\usepackage[caption=false]{subfig}
\usepackage[export]{adjustbox}
\usepackage{tikz}
\usepackage{slashed}
\usepackage{bm}
\usetikzlibrary{through,calc}
\usetikzlibrary{positioning}

\newcommand{\be}{\begin{equation}}
\newcommand{\ee}{\end{equation}}

\newcommand{\bea}{\begin{eqnarray}}
\newcommand{\eea}{\end{eqnarray}}

\newcommand{\bmt}{\left[\begin{matrix}}
\newcommand{\emt}{\end{matrix}\right]}

\begin{document}
\title{Strength of effective Coulomb interaction in two-dimensional  transition-metal Halides  \textit{MX$_2$}  and \emph{MX$_3$} (\emph{M}=Ti, V, Cr, Mn, Fe, Co, Ni; \emph{X}=Cl, Br, I)}

\author{Y. Yekta}
\affiliation{Department of Physics$,$ University of Guilan$,$ 41335-1914$,$ Rasht$,$ Iran}

\author{H. Hadipour}
\email{hanifhadipour@gmail.com}
\affiliation{Department of Physics$,$ University of Guilan$,$ 41335-1914$,$ Rasht$,$ Iran}

\author{E. \c{S}a\c{s}{\i}o\u{g}lu}
\email{ersoy.sasioglu@physik.uni-halle.de}
\affiliation{Institute of Physics$,$ Martin Luther University Halle-Wittenberg$,$ 06120 Halle (Saale) Germany}

\author{C. Friedrich}
\email{c.friedrich@fz-juelich.de}
\affiliation{Peter Gr\"unberg Institut and Institute for Advanced Simulation$,$ Forschungszentrum J\"ulich and JARA$,$ 52425 J\"ulich$,$ Germany}

\author{S. A. Jafari}
\email{jafari@sharif.edu}
\affiliation{Department of Physics$,$ Sharif University of Technology$,$ Tehran 11155-9161$,$ Iran}

\author{S. Bl\"{u}gel}
\affiliation{Peter Gr\"unberg Institut and Institute for Advanced Simulation$,$ Forschungszentrum J\"ulich and JARA$,$ 52425 J\"ulich$,$ Germany}

\author{I. Mertig}
\affiliation{Institute of Physics$,$ Martin Luther University Halle-Wittenberg$,$ 06120 Halle (Saale) Germany}

\date{\today}

\begin{abstract}
 We calculate the strength of the effective onsite Coulomb interaction (Hubbard $U$) in two-dimensional (2D) transition-metal 
(TM) dihalides \emph{MX$_2$} and trihalides \emph{MX$_3$} (\emph{M}=Ti, V, Cr, Mn, Fe, Co, Ni; \emph{X}=Cl, Br, I)
from first principles using the constrained random-phase approximation. 
The correlated subspaces are formed from $t_{2g}$ or $e_g$ bands at the Fermi energy. 
Elimination of the efficient screening taking place in these narrow bands gives rise to sizable interaction parameters $U$ between the localized $t_{2g}$ ($e_g$) electrons.
Due to 
this large Coulomb interaction, we find $U/W>$1 (with the band width $W$) in most TM halides,  making them strongly 
correlated materials. Among the metallic TM halides in paramagnetic state, the correlation strength $U/W$ reaches a 
maximum in Ni$X_{2}$ and Cr$X_{3}$ with values much larger than the corresponding values in elementary TMs 
and other TM compounds. Based on the Stoner model and the calculated $U$ and $J$ values, we discuss the tendency of the 
electron spins to order ferromagnetically.
\end{abstract}


\maketitle

\section{Introduction}
\label{sec1}

Since the discovery of graphene \cite{Geim,Katsnelson}
two-dimensional (2D) materials have been extensively studied due to their rich physical 
properties and diverse technological applications. One of the important applications is 
the use of such 2D systems in spintronics for logic and memory applications \cite{Kawakami,Ohno,Ashton}. 
In particular, 2D half-metallic magnets and spin-gapless semiconductors are desired 
for reconfigurable spintronic devices, which combines memory and logic into a single device 
\cite{Reconfigurable}. Despite substantial interest, most of the 2D materials are, however, 
not magnetic in their pristine form. From a theoretical point of view, according to the Mermin-Wagner 
theorem \cite{Mermin}, long-range magnetic order is not possible in 2D systems at finite 
temperatures, but this restriction is removed by magnetic anisotropy, which enables the formation 
of long-range magnetic order even in monolayers. Several standard approaches such as adsorption 
of atoms \cite{Sofo,Ersoy-0,Boukhvalov,Elias,Zhou}, point defects \cite{Hanif,Yazyev,Ugeda,McCreary,Nair}, 
and edge engineering \cite{Rao,Magda,Feldner,Li,Bagherpour} were developed to induce 
ferromagnetism in graphene and other graphene-like 2D materials. But these systematic 
ways are not well controlled for realistic applications. Therefore, 2D materials with 
intrinsic magnetism are of great interest for testing theories of magnetism in low dimensions 
as well as for ultralow-power memory and logic device applications.

Recently, intrinsic 2D ferromagnetism has been observed in materials containing 
transition-metal (TM) atoms such as in Cr$_2$Ge$_2$Te$_6$ \cite{Gong}, Fe$_3$GeTe$_2$, 
CrI$_3$ \cite{Huang,Kyle,Shcherbakov,Shabbir,Jiang}, VSe$_2$\cite{Bonilla}, and MnSe$_2$ \cite{OHara}.
For instance, it was observed that the CrI$_3$ monolayer exhibits ferromagnetic order 
below $T_c$=45 K \cite{Huang}. VSe$_2$ and MnSe$_2$ were reported to have itinerant 
ferromagnetic order even at room temperature \cite{Bonilla,OHara}. Indeed, the synthesis 
of the CrI$_3$ monolayer has led to a huge experimental and theoretical interest. 
First-principles theoretical studies have predicted that the long-range magnetic order 
is also possible in other 2D monolayers of TM dihalides and trihalides (\emph{MX$_2$} 
and \emph{MX$_3$}, \emph{M} = V, Cr, Mn, Fe and Ni, and \emph{X}=Cl, Br and I) \cite{McGuire,Ping,Tomar,Wang,Webster,Zhang-1,Zhang-2,McGuire-2,Botana,Junjie-1,Vadym}.
Besides CrI$_{3}$, ferromagnetic order in the other TM halides such as CrCl$_{3}$ \cite{McGuire-0}, 
CrBr$_{3}$ \cite{Zhaowe,Michael}, NiI$_{2}$ \cite{Kurumaji}, and VI$_{3}$  \cite{Kong,Son} 
monolayers were discovered experimentally and confirmed theoretically \cite{Torelli,Besbes,Tian,Junjie,Junyi}.

Due to the presence of narrow $t_{2g}$ or $e_g$ states at the Fermi level \cite{Wang} as
well as reduced screening and quantum confinement effects arising from reduced dimensionality, 
correlation effects are expected to play a crucial role in determining the electronic and 
magnetic properties of the 2D TM halides. Density functional theory (DFT) based 
on the local-spin-density approximation (LSDA) or generalized gradient approximation (GGA) 
may therefore not be a reliable method to calculate the physical properties of TM halides. In this respect, 
methods beyond DFT such as DFT+$U$ and DFT+DMFT might be necessary. Some TM halides have 
been studied by employing the DFT+$U$ method \cite{Torelli,Botana,Vadym}, in which the 
effective Coulomb interaction parameters $U$ are chosen arbitrarily or the $U$ values of the 
3D TMs are used. Only recently, a self-consistent constrained DFT method within linear response 
theory has been employed to calculate Hubbard $U$ parameters for VCl$_3$, VI$_3$ and CrX$_3$ 
(\emph{X}=Cl, Br, I) \cite{Junjie,Junyi}. The obtained $U$ parameters for V and Cr 3\textit{d} 
orbitals turn out to be close to the corresponding values in elementary transition metals and 
other TM compounds \cite{Ersoy,Yekta}.

The aim of this paper is a first-principles determination of the strength of the 
effective Coulomb interaction (Hubbard $U$) between localized electrons in 2D TM 
halides \emph{MX$_2$} and \emph{MX$_3$} (\emph{M} = V, Cr, Mn, Fe and Ni, and 
\emph{X} = Cl, Br and I) by employing the constrained random-phase approximation 
(cRPA) approach \cite{cRPA_1,cRPA_2,cRPA_3} within the full-potential linearized augmented 
plane-wave (FLAPW) method using maximally localized Wannier functions (MLWFs) \cite{Marzari,Freimuth}. 
We find the Hubbard $U$ parameters for $t_{2g}$ or $e_{g}$ electrons in metallic TM halides 
(paramagnetic state) vary between 1.0 and 5.1 eV, giving correlation strengths $U/W>1$, larger than corresponding values in elementary TMs and other TM compounds demonstrating strong electronic correlation in the TM halides. Furthermore, based on 
the Stoner model, we use the calculated $U$ and $J$ values to assess the stability of the 
ferromagnetic ordering. 

The rest of the paper is organized as follows: In Sec.\,\ref{sec2} we briefly present
the computational method and the cRPA method. In Sec.\,\ref{sec3}, we present calculated 
values of Coulomb interaction parameters for \emph{MX$_2$} and \emph{MX$_3$} for TM halides. 
Finally, we summarize our conclusions in Sec.\,\ref{sec4}.

\section{Computational method}
\label{sec2}

\begin{figure}[t]
\begin{center}
\vspace*{-0.3cm}
\includegraphics[scale=0.45]{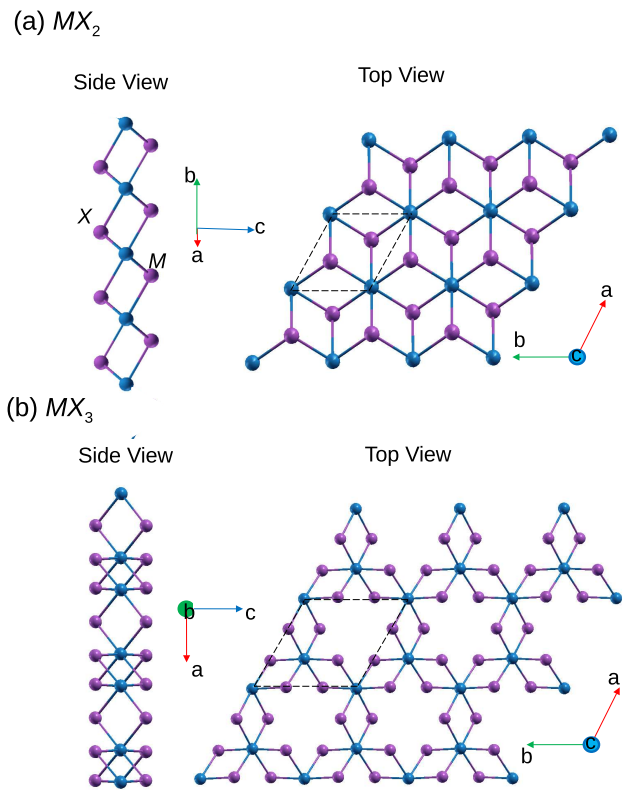}
\end{center}
\vspace*{-0.7cm} \caption{(Colors online) (a) Top and side views of the two-
dimensional crystal structure of TM dihalides \emph{MX$_2$} (b) Top and side 
views of the two-dimensional crystal structure of TM trihalides \emph{MX$_3$}. 
The blue and purple circles exhibit $M$ and $X$ atoms respectively. The $M$ atoms in (a) form a frustrated triangular lattice, while in (b) their honeycomb
structure is a bi-partite and hence non-frustrated lattice.} \label{fig1}
\end{figure}

We consider 2D TM halides with formulas \emph{MX$_{2}$} and \emph{MX$_{3}$}
(\emph{M}=Ti, V, Cr, Mn, Fe, Co, Ni; \emph{X}=Cl, Br, I). Fig.\,\ref{fig1}(a)  and 
Fig.\,\ref{fig1}(b) show the side and top view of crystal structures \emph{MX$_{2}$} 
dihalides and \emph{MX$_{3}$} trihalides, respectively. The lattice of TM dihalides 
consists of triangular nets of TM atoms and exhibits geometrical frustration when the 
magnetic moments couple antiferromagnetically. On the other hand, the TM atoms form 
honeycomb nets in \emph{MX$_{3}$} trihalide monolayers. The lattice parameters are
taken from Refs. \cite{Botana,McGuire-2,Tomar}. Simulation of $MX_2$ and $MX_3$ 
unit cells, containing one and two formula units, respectively, is based on the slab model 
having a 25 \AA{} vacuum separating them. For DFT calculations we use the FLEUR 
code \cite{fleur_code}, which is based on the FLAPW method. 
For the exchange correlation functional we use the generalized gradient approximation (GGA) 
parametrized by Perdew \emph{et al.} \cite{Perdew} (PBE). A 18$\times$18$\times$1 $\mathbf{k}$-point 
grid is used for all systems. A linear momentum cutoff of $G_{\mathrm max}$ = 4.5 bohr$^{-1}$ is 
chosen for the plane waves. The effective Coulomb interaction parameters are 
calculated within the cRPA method \cite{cRPA_1,cRPA_2,cRPA_3} implemented in the SPEX 
code \cite{Friedrich} with Wannier orbitals constructed from projection onto localized muffin-tin orbitals \cite{Freimuth}. A dense 16$\times$16$\times$1 k-point grid is used for the cRPA 
calculations.

\begin{figure}
\centering
\subfloat{%
  \includegraphics[width=85mm]{fig2a.eps}%
  \label{fig:subm4a}%
} \vspace{0.5cm}
\subfloat{%
  \includegraphics[width=85mm]{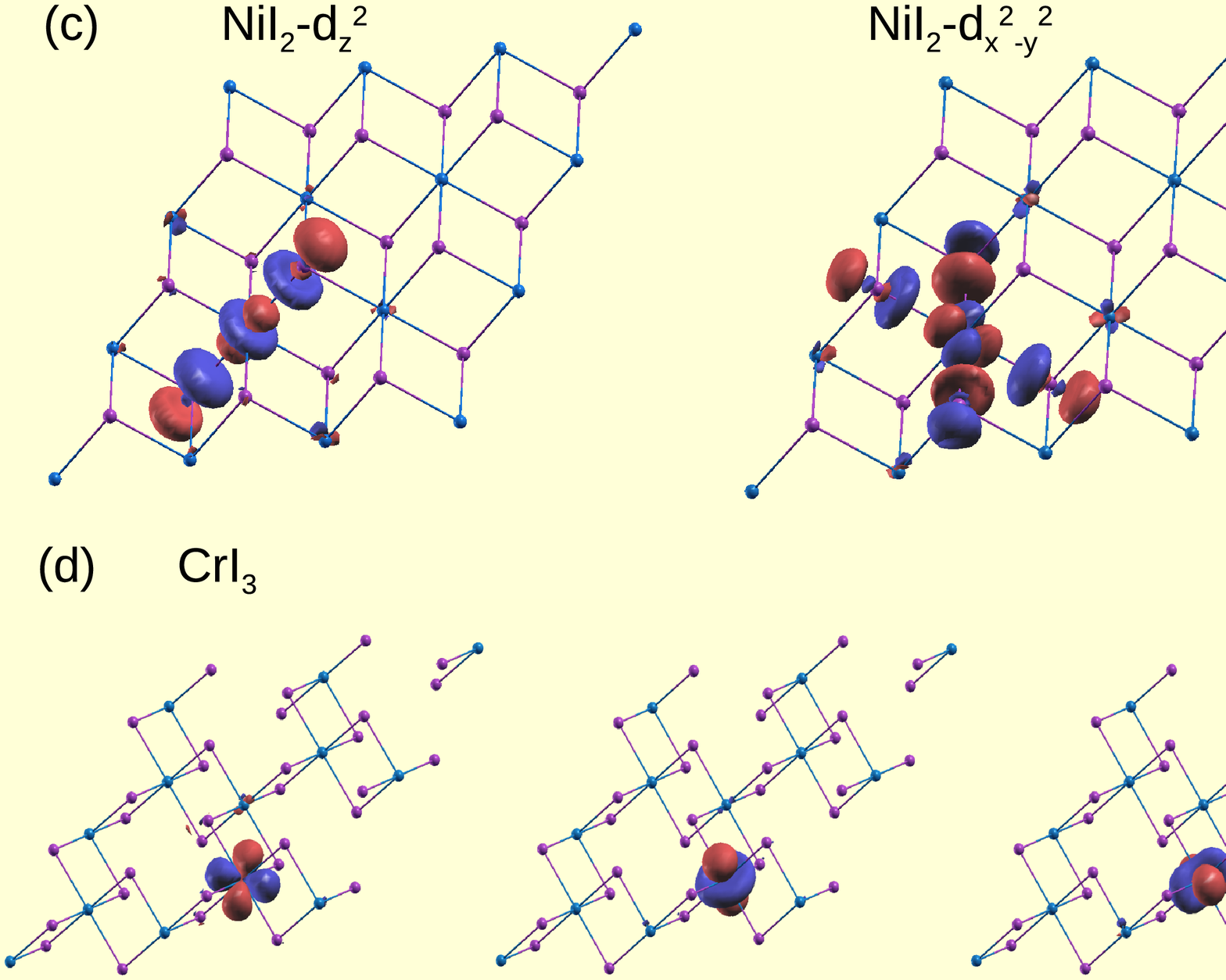}%
  \label{fig:subm4b}%
}
\vspace{-0.5cm}
\caption{(Color online) DFT-PBE (red) and Wannier-interpolated band
structures (blue) of non-spin-polarized (a) NiI$_2$ and (b) CrI$_3$. (c) The
$e_{g}$-like MLWFs for Ni atoms of NiI$_{2}$. (d) The
$t_{2g}$-like MLWFs for Cr atoms of CrI$_{3}$. }
\end{figure}

\begin{figure*}[t]
\begin{center}
\includegraphics[scale=0.34]{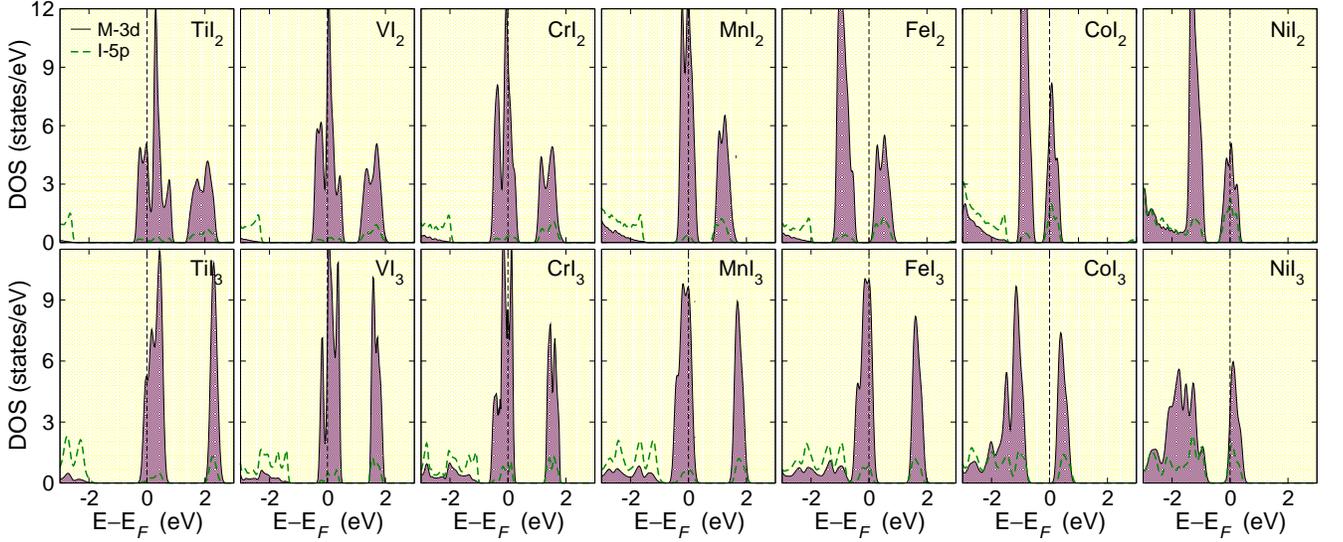}
\end{center}
\vspace*{-0.6cm} 
\caption{(Colors online) Orbital-resolved DOS for the non-spin-polarized $M$I$_2$ and $M$I$_3$ 
TM halides. Each panel shows the DOS projected onto $3d$ states of the \emph{M} 
atom as well as on $5p$ states of the I atom.  The two distinct groups of $d$ bands
correspond to $t_{2g}$ and $e_g$ bands, respectively.
The former overlaps with $5p$ states in the late TM iodides.} 
\label{fig3}
\end{figure*}

To identify the correlated subspace and construct Wannier functions properly, non-spin-polarized 
projected density of states (DOS)  are calculated for all systems and MLWFs are constructed for 
$t_{2g}$ or $e_g$ orbitals. To verify the validity of the calculated Wannier functions, in 
Fig.\,\ref{fig:subm4a}(a) and Fig.\,\ref{fig:subm4b}(b) we present a comparison of the non-spin-polarized 
DFT-PBE band structures with the corresponding Wannier-interpolated band structures obtained with 
$e_g$ and $t_{2g}$ Wannier orbitals for NiI$_{2}$ and CrI$_{3}$, respectively. In all cases, 
the original and the Wannier-interpolated bands agree very well. The bands around the Fermi energy, 
formed by Ni-$e_g$ (Cr-$t_{2g}$) orbitals in NiI$_2$ (CrI$_3$), are well separated from the rest 
of the bands. They can thus be employed to define an effective two-orbital (six-orbital) low-energy 
Hamiltonian. In a similar fashion, correlated subspaces can be defined for all considered systems, 
of $t_{2g}$ and $e_g$ character for early and late TM halides, respectively, as shown in Fig.\,\ref{fig3}.
We note that the $t_{2g}$-$e_{g}$ splitting is small for some systems (in particular, for some of the dihalides)\cite{Kvashnin}. It could, therefore, be necessary to go beyond the present minimal subspace by including, for example, the full $d$ shell or by considering spin polarization. However, it is already obvious from Fig.\,\ref{fig3} that one might then encounter the problem of entangled bands, which, on the one hand, complicates the Wannier construction and, on the other, makes the elimination of the subspace screening Eq.~(\ref{rpapol2}) less straightforward. For the present comparative study of the electron-electron interaction strength in a large class of materials, we therefore restrict ourselves to the minimal subspaces, all formed by isolated sets of bands, which will capture the essential physics of these materials. Our calculations may serve as reference point for more sophisticated studies of any of the investigated materials in the future.

Due to the systems' symmetry, the bands are not of pure $t_{2g}$ and $e_g$ character but are 
mixtures and also exhibit admixture from I-$p$ states. The denomination "$t_{2g}$" and "$e_g$"  
thus refers to their dominant orbital character. We describe the orbitals in more detail in 
the following.

In all systems, the TM atoms are each bound to six halogen atoms in octahedral coordination. The 
octahedron is, however, tilted with respect to the standard cartesian coordinate system, which has 
the $z$-axis perpendicular to the layers. Furthermore, the octahedron is distorted because of the 
two-dimensionality of the structure. Opposite halogens are still exactly opposite (forming a 
halogen-TM-halogen bond angle of 180$^\circ$), whereas of the remaining twelve halogen-TM-halogen 
bond angles six are slightly below and six are slightly above 90$^\circ$. For example, in NiI$_2$ (CrI$_3$) 
the difference to the right angle is 8.62$^\circ$ (1.13$^\circ$). 

The octahedron is tilted in such a way that two opposite faces of the octahedron are parallel to the layers.
If we take the three Ni-I bonds in NiI$_2$, whose iodine ends form the corners of one of these octahedron 
faces, as $x'$-, $y'$- and $z'$-axes, we can identify local $e_g$ orbitals in Fig.\,\ref{fig:subm4b}(c).
The orbitals show a strong admixture of iodine $p$ states, a delocalization effect which will be reflected 
in reduced interaction parameters later-on. Still, the $d_{{z'}^2}$ and $d_{{x'}^2-{y'}^2}$ orbital character 
is readily seen. In CrI$_3$, the octahedron is similarly tilted, and we can consider local $x'$-, $y'$-, 
$z'$-axes as above. While two of the orbitals presented in Fig.\,\ref{fig:subm4b}(d) indeed look like 
$t_{2g}$ orbitals, the one in the middle actually has the form of a $d_{z^2}$ orbital, which is, however, 
oriented along the cartesian $z$-axis, perpendicular to the layers. In fact, if we linearly combine the 
local $t_{2g}$ orbitals by $(d_{x'y'}+d_{y'z'}+d_{x'z'})/\sqrt{3}$, we obtain a $d_{z^2}$ orbital oriented 
perpendicular to the layer. This linear combination results from the breaking of octahedral symmetry 
caused by the layer structure. The other two $t_{2g}$ orbitals, $(d_{y'z'}-d_{x'y'})/\sqrt{2}$ and $(d_{x'y'}+d_{y'z'}-2d_{x'z'})/\sqrt{6}$, can be described, respectively, as a $d_{x'y'}$ rotated by 
45$^\circ$ around the $y$-axis and a distorted $d_{x'z'}$ orbital.

The fully screened Coulomb interaction $\tilde{U}$  
is related to the bare Coulomb interaction $V$ by
\begin{equation}
\tilde{U}(\boldsymbol{r},\boldsymbol{r}',\omega)=\int d\boldsymbol{r}''  
\epsilon^{-1}(\boldsymbol{r},\boldsymbol{r}'',\omega) V(\boldsymbol{r}'',\boldsymbol{r}'),
\label{fullysw}
\end{equation}
where $\epsilon(\boldsymbol{r},\boldsymbol{r}'',\omega)$ 
is the dielectric function. The dielectric function  is related to
the electron polarizability $P$ by
\begin{equation}
\epsilon(\boldsymbol{r},\boldsymbol{r}',\omega)=\delta(\boldsymbol{r}-\boldsymbol{r}')-\int d\boldsymbol{r}'' V(\boldsymbol{r},\boldsymbol{r}'')P(\boldsymbol{r}'',\boldsymbol{r}',\omega),
\label{rpadiel1}
\end{equation}
where the RPA polarization function $P(\boldsymbol{r}'',\boldsymbol{r}',\omega)$ is given by
\begin{equation}
\begin{gathered}
P(\boldsymbol{r},\boldsymbol{r}',\omega)= \\
2 \sum_{m}^{\mathrm occ} \sum_{m'}^{\mathrm unocc} \varphi_{m}(\boldsymbol{r}) \varphi_{m'}^{*}(\boldsymbol{r}) \varphi_{m}^{*}(\boldsymbol{r}') \varphi_{m'}(\boldsymbol{r}')  \\
\times\Bigg[ \frac{1}{\omega-\Delta_{mm'}+i\eta} - \frac{1}{\omega+\Delta_{mm'}-i\eta} \Bigg].
\end{gathered}
\label{rpapol1}
\end{equation}
Here, $\varphi_{m}(\boldsymbol{r})$ are the single-particle DFT Kohn-Sham eigenfunctions,
and $\eta$ a positive infinitesimal. $\Delta_{mm'}=\epsilon_{m'}-\epsilon_{m}$ with the Kohn-Sham eigenvalues
$\epsilon_{m}$. ß

In the cRPA approach, in order to exclude the screening due to the correlated subspace, we 
separate the full polarization function of Eq.~(\ref{rpapol1}) into two parts
\begin{equation}
P=P_{d}+P_{r},
\label{rpapol2}
\end{equation}
where $P_{d}$ includes only the transitions ($m\rightarrow m'$) between the states of the correlated subspace
and $P_{r}$ is the remainder. Then, the frequency-dependent effective Coulomb interaction 
is given schematically by the matrix equation
\begin{equation}
U(\omega) = [1-VP_{r}(\omega)]^{-1}V.
\end{equation}
It contains, in $P_r$, screening processes that would not be
captured by the correlated subspace and excludes the ones that
take place within the subspace.

The matrix elements of the effective Coulomb interaction in the MLWF basis are 
given by
\begin{equation}
\begin{gathered}
U_{\mathbf{R}n_{1},n_{3},n_{2},n_{4}}(\omega)= \\
\int\int d\boldsymbol{r}d\boldsymbol{r}' w_{n_{1}\mathbf{R}}^{*}(\boldsymbol{r}) w_{n_{3}\mathbf{R}}(\boldsymbol{r}) U(\boldsymbol{r},\boldsymbol{r}',\omega) w_{n_{4}\mathbf{R}}^{*}(\boldsymbol{r}') w_{n_{2}\mathbf{R}}(\boldsymbol{r}'),
\end{gathered}
\label{hubudef31}
\end{equation}
where $w_{n\mathbf{R}}(\boldsymbol{r})$ is the MLWF at site $\mathbf{R}$ with orbital index
$n$, and the effective Coulomb potential $U(\boldsymbol{r},\boldsymbol{r}',\omega)$ is calculated 
within the cRPA as described above. We define the average Coulomb matrix elements  $U$, $U'$,
and $J$ in the static limit ($\omega=0$) as follows \cite{Anisimov1,Anisimov2}:
\begin{equation}
U=\frac{1}{L}\sum_{m}
U_{mm;mm}\,,
\label{U_diag}
\end{equation}
\begin{equation}
U'=\frac{1}{L(L-1)}\sum_{m \neq n}
U_{mn;mn}\,,
\label{U_offdiag}
\end{equation}
\begin{equation}
J=\frac{1}{L(L-1)}\sum_{m \neq n}U_{mn;nm}\,,
\label{Hund_J}
\end{equation}
where $L$ is the number of localized orbitals, i.e., two for $e_{g}$ and three for $t_{2g}$ orbitals.
This parametrization of partially screened Coulomb interactions is the so-called Hubbard-Kanamori 
parametrization. Similar to the definition of $U$ ($U'$, $J$), we can also define the so-called 
fully screened interaction parameters $\tilde{U}$  ($\tilde{U'}$, $\tilde{J}$)  as well 
as unscreened (bare) $V$. The bare $V$ provides information about the localization of Wannier functions
and is a useful parameter in the interpretation of the screened Coulomb interaction parameters.

\section{Results and discussion}
\label{sec3}

In the low-energy model Hamiltonian description of correlated solids, the non-interacting 
one-electron part of the effective model is defined for a system, in which there is 
no spontaneous symmetry breaking, i.e., it is a paramagnetic (non-spin-polarized) metal. 
The calculation of effective Coulomb interaction parameters (Hubbard $U$) should therefore
be based on such a system.

To identify the correlated subspace for all studied TM halides, we present in  Fig.\,\ref{fig3} 
orbital-resolved DOS for $M$I$_{2}$ and $M$I$_{3}$. In all compounds except for the semiconducting  
FeI$_2$ and CoI$_3$, the Fermi energy falls into a group of bands, which are of $t_{2g}$ character
for the early halides, $M=$ Ti to Mn (Ti to Fe) for $M$I$_2$ ($M$I$_3$), and of $e_g$ character 
for the late halides, $M=$ Co and Ni (Ni) for $M$I$_2$ ($M$I$_3$). These bands are assumed to 
form minimal correlated subspaces in this work.
On the other hand, FeI$_2$ and CoI$_3$ are semiconducting 
with the Fermi level falling in the energy gap between the $t_{2g}$ and $e_{g}$ bands. 
Depending on the type of doping, electron or hole doping, only one type
of bands will form the minimal correlated subspace. For these systems, we will
present Hubbard $U$ parameters for $t_{2g}$, $e_g$, and $d$ orbitals. 
The first two correspond to the zero-doping limit (or very dilute doping), since the states of the subspace are either all occupied or all empty. 
As hence no screening takes place in the subspace, the partially and fully screened parameters (e.g., $U$ and $\tilde{U}$) are identical.
The orbital-resolved DOS of $MX_2$ and $MX_3$ with $X=$ Cl and Br look 
very similar, so the subspaces can be defined identically to those of the iodides. 
The $U$ matrix element of the p-admixed $t_{2g}$ state differs from the pure $t_{2g}$ state by maximally 0.1 eV among the materials. The calculated Hubbard $U$ values should therefore be applicable to standard DFT+$U$ implementations that are based on atomic bases, as well, not only to implementations that employ Wannier functions.

In Table\,\ref{table1}, we present the onsite average intra-orbital unscreened 
(bare) Coulomb interaction  $V$, partially (fully) screened $U$ ($\tilde{U}$), as well as average 
inter-orbital $U'$ ($\tilde{U'}$) and exchange parameter $J$ ($\tilde{J}$). The behavior of the 
bare interaction $V$ for $t_{2g}$ orbitals across the 3$d$ TM atoms, from Ti to Fe, in $MX_2$ is similar 
to the case of elementary TMs. The $V$ parameter increases nearly linearly with increasing electron 
number, which is due to the contraction of the wave functions with increased nuclear charge and the concomitant increased localization of the Wannier functions.
By contrast, the $e_g$ orbitals exhibit the opposite trend from Fe to Ni in $MX_2$: The $V$ decreases.
An analysis of the shape of the Wannier orbitals reveals that the coupling to neighboring halogen p states gets stronger, which makes the orbitals increasingly spill into these states giving rise to a delocalization and, therefore, to smaller $V$ parameters.
The same trend is seen for $MX_3$ from Co to Ni.
The behavior of $V$ for the $t_{2g}$ orbitals in the trihalides is somewhat different from that of the dihalides: While early in the series from Ti to Co we see a similar increase as in the dihalides, the values go over a maximum and drop off sharply for Co$X_3$.
In all cases, we see a decrease of the bare $V$ for the halide series $MX_{2}$ with $X=$ Cl to I, which is likely caused by the increase of the lattice constant in this order, making the orbitals more extended.

The effects are reflected also in the band widths $W$ of the $t_{2g}$ and $e_g$ bands presented in Table\,\ref{table1} 
(also see Fig.\,\ref{fig3}), which tend to decrease from Ti to Ni, similar to the case of 
elementary $3d$ TMs with the difference that, due to reduced coordination (reduced hybridization) 
in 2D, the $W$ is much smaller than the corresponding values in $3d$ TMs, making the $t_{2g}$ 
and  $e_{g}$ peaks in the DOS sharper and more atomic-like.

\begin{table*}[htbp]
\caption{Lattice constants $a$, orbital type of correlated
subspace, bandwidth $W$,  onsite average intra-orbital bare $V$, partially (fully) screened  $U$ ($\tilde{U}$), 
inter-orbital partially (fully) screened $U'$ ($\tilde{U}'$), partially (fully) screened exchange interaction 
$J$ ($\tilde{J}$), correlation strength $U/W$, and the DOS at the Fermi level $D$(E$_F$)  for $MX_2$ and
$MX_3$ compounds.}
\centering
\begin{ruledtabular}
\begin{tabular}{lccccccccc}
 & \multicolumn{9}{c}{}\\
 \emph{MX$_{2/3}$}  & a (\AA) & orbitals& $W$(eV)  & $V$(eV)&$U$  ($\tilde{U}$) (eV)& $U'$ ($\tilde{U}'$) (eV)  & $J$($\tilde{J}$) (eV) &$U/W$ & $D$(E$_F$)\\ \hline
TiCl$_2$ & 3.56 & $t_{2g}$ & 1.43 & 14.07 & 5.07 (0.87)  &  4.22 (0.43) &  0.43 (0.23) & 3.55 & 0.45  \\
TiBr$_2$ & 3.73 & $t_{2g}$ & 1.31 & 13.32 & 4.26 (0.79)  &  3.47 (0.41) &  0.39 (0.19) & 3.25 & 0.51  \\
TiI$_2$  & 4.11 & $t_{2g}$ & 1.15 & 11.70 & 3.26 (0.63)  &  2.61 (0.32) &  0.33 (0.16) & 2.83 & 0.73  \\
VCl$_2$  & 3.62 & $t_{2g}$ & 1.20 & 15.95 & 4.60 (0.95)  &  3.62 (0.63) &  0.51 (0.16) & 3.83 & 1.28  \\
VBr$_2$  & 3.81 & $t_{2g}$ & 1.03 & 14.96 & 3.98 (0.86)  &  3.03 (0.57) &  0.49 (0.16) & 3.86 & 1.91  \\
VI$_2$   & 4.08 & $t_{2g}$ & 0.94 & 13.63 & 3.10 (0.44)  &  2.33 (0.27) &  0.39 (0.10) & 3.30 & 2.25  \\
CrCl$_2$ & 3.55 & $t_{2g}$ & 1.05 & 17.32 & 4.13 (0.23)  & 3.13 (0.03)  &  0.54 (0.09) & 3.93 & 4.50  \\
CrBr$_2$ & 3.74 & $t_{2g}$ & 0.99 & 16.51 & 3.64 (0.23)  &  2.68 (0.06) &  0.51 (0.08) & 3.68 & 4.87  \\
CrI$_2$  & 3.99 & $t_{2g}$ & 0.96 & 14.88 & 2.96 (0.26)  &  2.16 (0.07) &  0.42 (0.10) & 3.08 & 3.64  \\
MnCl$_2$ & 3.64 & $t_{2g}$ & 0.85 & 19.10 & 3.71 (0.82)  &  2.69 (0.62) &  0.56 (0.08) & 4.36 & 6.41  \\
MnBr$_2$ & 3.84 & $t_{2g}$ & 0.81 & 18.12 & 3.25 (0.29)  &  2.29 (0.06) &  0.52 (0.07) & 4.01 & 6.02  \\
MnI$_2$  & 4.12 & $t_{2g}$ & 0.75 & 17.19 & 2.76 (0.26)  &  1.90 (0.04) &  0.46 (0.06) & 3.68 & 4.18  \\
FeCl$_2$ & 3.49 & $t_{2g}$ & 1.10 & 20.11 & 3.50 (3.50)  &  2.43 (2.43) &  0.56 (0.56) & 3.18 & 0.00  \\
         &      & $e_{g}$  & 0.70 & 17.00 & 3.06 (3.06)  & 2.06 (2.06)  &  0.50 (0.50) & 4.37 & 0.00  \\
         &      & $d$      & 2.41 & 19.96 & 5.97 (3.21)  & 4.84 (2.42)  &  0.57 (0.42) & 2.48 & 0.00  \\
FeBr$_2$ & 3.69 & $t_{2g}$ & 0.97 & 19.02 & 3.14 (3.14)  & 2.11 (2.11)  &  0.53 (0.53) & 3.24 & 0.00  \\
         &      & $e_{g}$  & 0.78 & 15.24 & 2.64 (2.64)  & 1.76 (1.76)  &  0.44 (0.44) & 3.38 & 0.00  \\
         &      & $d$      & 2.37 & 19.13 & 5.21 (3.07)  & 3.86 (2.20)  &  0.55 (0.42) & 2.20 & 0.00  \\
FeI$_2$  & 3.98 & $t_{2g}$ & 0.80 & 17.39 & 2.52 (2.52)  & 1.66 (1.66)  &  0.44 (0.44) & 3.15 & 0.00  \\
         &      & $e_{g}$  & 0.91 & 12.83 & 2.03 (2.03)  & 1.36 (1.36)  &  0.33 (0.33) & 2.23 & 0.00  \\
         &      & $d$      & 2.21 & 17.64 & 4.07 (3.05)  & 2.98 (1.67)  &  0.53 (0.38) & 1.84 & 0.00  \\
CoCl$_2$ & 3.49 & $e_{g}$  & 0.59 & 16.51 & 3.18 (0.38)  & 2.20 (0.26)  &  0.49 (0.06) & 5.39 & 3.26  \\
CoBr$_2$ & 3.73 & $e_{g}$  & 0.61 & 14.93 & 2.67 (0.19)  & 1.82 (0.07)  &  0.42 (0.06) & 4.38 & 3.51  \\
CoI$_2$  & 3.92 & $e_{g}$  & 0.64 & 12.26 & 2.16 (0.16)  & 1.52 (0.03)  &  0.32 (0.07) & 3.38 & 4.54  \\
NiCl$_2$ & 3.45 & $e_{g}$  & 0.53 & 15.17 & 3.37 (0.16)  & 2.43 (0.04)  &  0.47 (0.06) & 6.36 & 4.12  \\
NiBr$_2$ & 3.64 & $e_{g}$  & 0.59 & 13.52 & 3.09 (0.22)  & 2.27 (0.07)  &  0.41 (0.07) & 5.24 & 3.28  \\
NiI$_2$  & 3.94 & $e_{g}$  & 0.62 & 10.60 & 2.36 (0.44)  & 1.78 (0.27)  &  0.29 (0.09) & 3.81 & 2.76  \\ \\
TiCl$_3$ & 5.88 & $t_{2g}$ & 1.16 & 14.48 & 4.62 (0.68)  &  3.75 (0.24)  & 0.46 (0.22) & 3.98 & 1.09 \\
TiBr$_3$ & 6.27 & $t_{2g}$ & 1.08 & 14.15 & 3.97 (0.51)  &  3.12 (0.16)  & 0.43 (0.17) & 3.68 & 0.82 \\
TiI$_3$  & 6.62 & $t_{2g}$ & 1.01 & 12.86 & 2.99 (0.37)  &  2.26 (0.06)  & 0.38 (0.15) & 2.96 & 0.61 \\
VCl$_3$  & 6.04 & $t_{2g}$ & 0.91 & 16.28 & 4.49 (0.31)  &  3.46 (0.04)  & 0.54 (0.14) & 4.93 & 1.17 \\
VBr$_3$  & 6.33 & $t_{2g}$ & 0.86 & 15.59 & 3.82 (0.36)  &  2.84 (0.09)  & 0.51 (0.14) & 4.44 & 1.86 \\
VI$_3$   & 6.86 & $t_{2g}$ & 0.82 & 14.47 & 2.91 (0.28)  &  2.07 (0.06)  & 0.44 (0.12) & 3.55 & 2.44 \\
CrCl$_3$ & 5.75 & $t_{2g}$ & 0.81 & 16.53 & 4.08 (0.48)  &  3.07 (0.20)  & 0.53 (0.17) & 5.04 & 2.72 \\
CrBr$_3$ & 6.34 & $t_{2g}$ & 0.76 & 16.24 & 3.50 (0.31)  &  2.52 (0.12)  & 0.51 (0.10) & 4.61 & 2.40 \\
CrI$_3$  & 6.85 & $t_{2g}$ & 0.61 & 15.29 & 2.67 (0.25)  &  1.84 (0.10)  & 0.44 (0.08) & 4.38 & 2.47 \\
MnCl$_3$ & 6.08 & $t_{2g}$ & 0.85 & 18.05 & 3.64 (0.56)  &  2.63 (0.30)  & 0.55 (0.13) & 4.28 & 2.15 \\
MnBr$_3$ & 6.39 & $t_{2g}$ & 0.79 & 17.28 & 3.15 (0.55)  &  2.19 (0.31)  & 0.51 (0.13) & 3.99 & 3.19 \\
MnI$_3$  & 6.85 & $t_{2g}$ & 0.65 & 15.73 & 2.62 (0.48)  &  1.61 (0.26)  & 0.44 (0.12) & 4.03 & 3.39 \\
FeCl$_3$ & 6.05 & $t_{2g}$ & 0.82 & 18.14 & 3.30 (0.33)  &  2.32 (0.14)  & 0.54 (0.09) & 4.02 & 2.52 \\
FeBr$_3$ & 6.43 & $t_{2g}$ & 0.78 & 17.24 & 2.85 (0.29)  &  1.92 (0.11)  & 0.50 (0.09) & 3.65 & 2.47 \\
FeI$_3$  & 6.97 & $t_{2g}$ & 0.69 & 15.29 & 2.14 (0.22)  &  1.37 (0.03)  & 0.41 (0.08) & 3.10 & 2.62 \\
CoCl$_3$ & 6.07 & $t_{2g}$ & 0.65 & 16.82 & 2.95 (2.95)  &  2.05 (2.05)  & 0.50 (0.50) & 4.54 & 0.00 \\
         &      & $e_{g}$  & 0.45 & 13.95 & 2.43 (2.43)  &  1.67 (1.67)  & 0.38 (0.38) & 5.40 & 0.00 \\
         &      & $d$      & 1.72 & 16.21 & 3.60 (2.70)  &  2.67 (2.04)  & 0.47 (0.35) & 2.09 & 0.00 \\
CoBr$_3$ & 6.30 & $t_{2g}$ & 0.93 & 14.47 & 2.45 (2.45)  &  1.67 (1.67)  & 0.43 (0.43) & 2.63 & 0.00 \\
         &      & $e_{g}$  & 0.78 & 12.84 & 2.11 (2.11)  &  1.42 (1.42)  & 0.34 (0.34) & 2.71 & 0.00 \\
         &      & $d$      & 2.45 & 14.27 & 3.08 (2.23)  &  2.26 (1.71)  & 0.39 (0.31) & 1.26 & 0.00 \\
CoI$_3$  & 6.81 & $t_{2g}$ & 1.62 & 10.34 & 1.34 (1.34)  &  0.94 (0.94)  & 0.24 (0.24) & 0.83 & 0.00 \\
         &      & $e_{g}$  & 1.10 & 10.92 & 1.48 (1.48)  &  0.96 (0.96)  & 0.26 (0.26) & 1.35 & 0.00 \\
         &      & $d$      & 3.32 & 10.65 & 2.10 (1.40)  &  1.55 (1.05)  & 0.24 (0.18) & 0.63 & 0.00 \\
NiCl$_3$ & 6.05 & $e_{g}$  & 0.63 & 11.76 & 2.01 (0.15)  &  1.39 (0.04)  & 0.31 (0.06) & 3.19 & 2.52 \\
NiBr$_3$ & 6.16 & $e_{g}$  & 0.75 & 10.55 & 1.79 (0.25)  &  1.22 (0.11)  & 0.28 (0.07) & 2.39 & 1.85 \\
NiI$_3$  & 6.64 & $e_{g}$  & 0.91 & 8.55  & 1.08 (0.24)  &  0.69 (0.10)  & 0.20 (0.07) & 1.19 & 1.40 \\

\end{tabular}
\label{table1}
\end{ruledtabular}
\end{table*}

To discuss the partially screened (Hubbard $U$) effective Coulomb interaction parameters, we
focus on the TM iodides. As seen in  Table\,\ref{table1}, the $U$ values for $M$ sites in 
$M$I$_{2}$ ($M$I$_{3}$) compounds vary between 2.1$-$3.3 eV (1.1$-$3.0 eV) and decrease with 
moving from Ti to Ni, which can be described by the projected density of states in Fig.\,\ref{fig3}.
Just below the $d$ states there is a broad peak of iodine $5p$ states, which should contribute with
$5p\rightarrow d$ transitions sizably to the screening. Across the series Ti to Ni, the 
$5p$ states are seen to approach the $d$ states, which effectively increases the electronic screening and, thus, 
acts to compensate the increase of $U$ caused by Wannier localization, giving rise to the reduction of $U$
parameters with increasing $3d$ electron number in both types of TM halides. The same behaviour is 
observed in Br- and Cl-based TM halides (see Table.\,\ref{table1}).

Moving upwards in the group of halogens, from I to Cl, the \emph{M}$-$\emph{X} bond lengths decrease. 
As a consequence, orbitals overlap more strongly and hybridization increases, pushing 
the states apart energetically. The larger energy difference makes electronic transitions $p\rightarrow t_{2g}$
less likely and reduces electronic screening. From this, one would expect to see
a tendency to larger Hubbard $U$ values, which is, in fact, what we observe in Table
\ref{table1}.



Comparing the partially screened ($U$) with the fully screened parameters ($\tilde{U}$)  
gives information about the screening within the correlated subspace. 
Except for the semiconductors Fe$M_2$ and Co$M_3$, this screening is metallic. 
It is very efficient and becomes the dominant screening channel.
As a consequence, the fully screened parameters $\tilde{U}$ show a behavior very different from $U$: The values are all very small for the metallic systems and fall in the range between 0.1 and 0.5 eV, about one fifth of the $U$ values. They do not follow a specific trend across the TM series, nor do they follow a general ordering with respect to the different kinds of halides (Cl, Br, I) as in the partially screened $U$ parameters.
Our results show that $t_{2g}\rightarrow t_{2g}$ and $e_{g}\rightarrow e_{g}$ transitions 
 contribute substantially to the screening of the fully screened Coulomb interaction $\tilde{U}$ in metallic 
systems, while in the semiconducting ones this screening vanishes, making $\tilde{U}$ identical to $U$ for the $t_{2g}$ and $e_g$ orbitals, whereas the $t_{2g}\rightarrow e_{g}$ transitions reduce $\tilde{U}$ with respect to $U$ for the full $3d$ shell.

In the semiconducting cases Fe$X_{2}$ and Co$X_{3}$, the $3d$ parameters show a behavior parallel 
to that of the $t_{2g}$/$e_g$ parameters and are larger by about 20-70\%, elucidating that 
transitions between $t_{2g}$ and $e_g$ states play an important quantitative role in the electronic 
screening but do not affect the screening qualitatively. 

In cubic symmetry, the Hubbard-Kanamori inter-orbital Coulomb interaction term $U'$ satisfies 
the relation $U'=U-2J$. This relation is nearly fulfilled in most TM halides, even though 
cubic symmetry is broken.
The $J$ parameters vary in the range 0.20$-$0.57 eV 
and show a behavior very much in parallel to the bare parameters $V$, despite the very different range of values, which reveals a more quantitative than qualitative effect of the electronic screening on the exchange parameters $J$.

We now compare our calculated Hubbard $U$ values with reported ones in the 
literature.  Besbes et al.
\cite{Besbes}
calculated  Hubbard $U$ values for bulk CrCl$_3$ and 
CrI$_3$ and 
 obtained $U$ values of 
1.79 eV (CrCl$_3$) and 1.15 eV (CrI$_3$), which are significantly smaller 
than our calculated Coulomb matrix elements presented  in Table\,\ref{table1}, 
whereas their exchange parameters are larger (0.85 and 0.78 eV) than our values. 
A possible reason for the disagreement is the different dimensionality of the systems, 3D versus 2D. 
3D materials usually offer more screening with the consequence of smaller $U$ parameters.
There are other possible reasons: Besbes et al. employ a  method different to ours,
a combined cLDA/RPA scheme \cite{Solovyev}.
They also used a different Wannier basis. They combined chromium $d$ and halogen $p$ states
into a larger Wannier basis for a $dp$ tight-binding description, whereas
we have employed minimal $t_{2g}$ and $e_g$ subspaces.
And, finally, the definition of the Hubbard $U$ parameter is different. 
We have calculated Hubbard-Kanamori parameters for the $t_{2g}$ and $e_g$ subsets (see Eq. (\ref{U_diag})), whereas Besbes et al. 
defined the Hubbard parameters for the full
atomic $p$ and $d$ shells. In this different kind of $U$ parameter the averaging is over the full matrix instead of just over the diagonal elements.


In the case of V-based trihalides, He et al. calculated $U$ values for VCl$_3$ 
and VI$_3$  using the self-consistent linear response method within the cLDA 
approach and obtained  $U$=3.35 eV (for VCl$_3$) and  3.68 eV (for VI$_3$) \cite{Junjie}. 
These values are closer to our results. 



In the following, we discuss the appearance of ferromagnetism in TM halides.
The ferromagnetic state is the ground state for most of the TM trihalides $MX_{3}$, 
while TM dihalides $MX_{2}$ exhibit diverse magnetic behavior ranging from 
half-metallic ferromagnetism (FeCl$_{2}$) to antiferromagnetism (VI$_{2}$) 
and from 120$^{\circ}$-antiferromagnetism (MnI$_{2}$) to helical magnetism 
(NiI$_{2}$). A  richness of magnetic phases that results from the 
triangular lattice of $M$ atoms, which frustrates the exchange coupling. Such a frustration
is not present for $MX_3$ compounds, where $M$ atoms reside on a bi-partite (honeycomb) lattice. The calculation of $U$ 
parameters for such complex magnetic ground states is beyond the scope of the 
present paper. Note that such a calculation would require the full $3d$ shell to be included in the correlated subspace. The present values
for the minimal $t_{2g}$ and $e_{g}$ subspaces would not be immediately applicable in this case.

\begin{figure}[t]
\begin{center}
\includegraphics[scale=0.325]{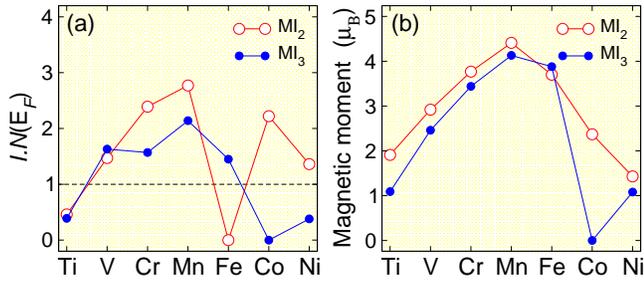}
\end{center}
\vspace*{-0.5cm} \caption{(Colors online)
(a) Stoner criterion for  $M$I$_2$ and $M$I$_3$ TM halides. (b) Calculated magnetic 
moments (in units of $\mu_{B}$) of TM atoms for  $M$I$_2$ and $M$I$_3$ TM halides.} \label{fig6}
\end{figure}

Among the TM trihalides, CrI$_{3}$ is of particular interest 
for spintronic applications as well as theories of low-dimensional magnetism 
because it was one of the first materials in which ferromagnetism was detected 
experimentally in the monolayer limit. It is a ferromagnetic semiconductor with 
a Curie temperature of 45 K in the monolayer and 61 K in the bulk. 
As the $MX_2$ and $MX_3$ halides contain partially filled $3d$ TM atoms, we can use the simple Stoner
model to discuss the appearance of ferromagnetism in these materials. The Stoner criterion for
ferromagnetism is given by  $I \cdot D$(E$_{F}$)$>$ 1, where $I$ is the Stoner parameter and $D$(E$_{F}$) 
is the DOS at the Fermi energy in the nonmagnetic state. The Hartree-Fock solution of the multiorbital 
Hubbard model gives a relationship between the Stoner parameter $I$, Hubbard $U$, and exchange $J$ 
by $I = (U + 6J )/5$\cite{Stollhoff}. Stollhoff et al. showed that in elementary TMs the electron 
correlations reduce $I$ by roughly 40$\%$.
We note that the magnetism in some of the TM halides might not be describable by the simple Stoner mechanism. For example, MnI$_2$ has an antiferromagnetic (120 degrees) ground state. For reasons of consistency, we nevertheless discuss the Stoner condition for all materials and leave more detailed analyses taking into account Hund’s rule coupling for future studies.
In Fig.\,\ref{fig6} (a)  we present  the  correspondingly scaled Stoner parameter 
$I \cdot D$(E$_F$) for iodine-based compounds. The experimentally observed and theoretically predicted 
ferromagnetic TM halides satisfy the Stoner criterion, and the paramagnetic state is unstable toward 
the formation of ferromagnetism, which reasonably agrees with the results of spin-polarized DFT total 
energy calculations and the large magnetic moments presented in Fig.\,\ref{fig6}(b).
The failure of the Stoner criterion to predict the ferromagnetism of  Fe$I_2$ is due to the fact that
this compound is a semiconductor, while the Stoner criterion is based on a metallic parent (non-interacting) limit.

With this tendency to form magnetic ground states, one might wonder whether and to what degree the Hubbard $U$ parameters would change if a magnetic ground state were taken as reference system instead of the non-spin-polarized one. For instance for the case of MnI$_2$, we obtain a $U$ value of  3.37 eV, which is not too different from the paramagnetic case (2.76 eV). This might be due to the fact that the formation of the magnetic moments predominantly affects the low-energy states, but these are just the ones that make up the correlated subspace whose screening channels are eliminated from the effective Hubbard $U$ interaction.

Finally, we want to briefly discuss the strength of the electronic correlations in 2D TM halides. 
Qualitatively, the correlation strength is defined as the ratio of the effective Coulomb interaction 
$U$ to the bandwidth $W$ ($U/W$). In Table\,\ref{table1} we present $U/W$ values for all TM halides. 
Note that the $U/W$ values are calculated for a non-spin-polarized (paramagnetic) state. In the case 
of $M$X$_2$ and ignoring the non-metallic system (Fe\emph{X$_{2}$}), the correlation strength $U/W$  tends to increase
from Ti- to Ni-based materials. 
 There is no clear trend in the trihalides, although a strong decrease of $U/W$ is noted for the $e_g$ subspace in the iodides and, less pronounced, also in the bromides.
For  most of the metallic TM halides, we find $U/W >$ 2 with
maxima for NiCl$_2$ and CrCl$_3$. We thus expect electron correlations to be strong 
in these materials. They should play an important role in model Hamiltonian studies of the TM halides.

 As a consequence of $U/W>1$, one may expect rich correlation phenomena such as 
magnetic order, Mott-insulating phases, etc. For instance, in contrast to experimental results\cite{Kong,Son} 
showing insulating behavior in VI$_{3}$, the partially filled bands in PBE calculations give  rise to a 
half-metallic behavior\cite{Tian,Junjie}. This implies that electron-electron interactions play a 
crucial role in electronic and magnetic properties of TM halides, especially in the systems with 
nearly half-filled $3d$ bands. Applying the DFT+$U$ method with $U$=3.8 eV to Mn\emph{X}$_{2}$ 
dihalides not only increases the band gap but also results in ferromagnetic order, whereas these 
systems remain antiferromagnetic when calculated without $U$ \cite{Vadym}. In the case of 
VI$_{3}$, DFT+$U$ employing $U$=3.5 eV
 opens a finite band-gap of about 0.84 eV \cite{Tian}, which is in agreement with experiments, 
while the system is a half-metal within DFT-PBE. VI$_3$ is therefore commonly classified as a Mott insulator. 
This inconsistency between experiment and DFT-PBE has also been  found for Cr\emph{X}$_{3}$, 
manifesting a possible Mott insulating state\cite{McGuire,Zhang-1}. Experimentally, CrI$_{3}$ is 
insulating\cite{Huang,Jiang,Kyle} even above the Curie temperature, which suggests that the band-gap 
does not stem from the exchange splitting (i.e, from magnetism) but
that the strong electron-electron correlation is 
responsible for the formation of the band gap in this material. There are theoretical works that argue that 
the band gap of Cr-based trihalides can be described  as a mixture of Mott-Hubbard and charge-transfer type\cite{Zhang-1}.
Note that layered materials that exhibit a Mott-insulating character are very rare. The Mott phase has 
been experimentally discovered in bilayer twisted graphene\cite{Shen-x,Seo-x,Choi-1}, $\sqrt{13} \times \sqrt{13}$ 
supercell of 1T-TaS(Se)$_{2}$ and 1T-NbSe$_{2}$ materials (the Star of David 
cluster)\cite{Ma-x,Perfetti,Ritschel}, and TM phosphorous trichalcogenides\cite{Kim1}, 
\emph{M}P$X_{3}$, where \emph{M} is a TM and \emph{X} are chalcogen elements, which were 
confirmed by \emph{ab initio} calculations \cite{Kim2,Lee-2,Choi-1}.
In all mentioned monolayers, the observed insulating phase has been discussed to originate from the presence of 
narrow bands around the Fermi energy. 
For example, in the distorted phase of 1T-TaS$_{2}$, there is a flat band with mainly $d_{z^{2}}$ character at $E_{\mathrm F}$ which increases $U/W$ substantially \cite{Tresca,Kamil,Darancet},
 even though the Coulomb interaction parameter $U$ has been calculated to be 0.4 eV only \cite{Kamil}, which is significantly 
smaller than the corresponding value $U$=2.27 eV in undistorted 1T-TaS$_{2}$ \cite{Cococcioni}.
Such narrow bands, of $t_{2g}$ or $e_g$ character, are also present in CrI$_{3}$, 
VI$_{3}$, and NiI$_{2}$,  resulting in a large $U/W$ difference between TM halides and elementary TMs.

\section{Summary and Outlook}
\label{sec4}

We have performed systematic \emph{ab initio} calculations to determine the strength of the 
effective Coulomb interaction (Hubbard $U$) between localized electrons in various 2D TM Halides 
with formulas \emph{MX$_{2}$} and \emph{MX$_{3}$} (\emph{M}=Ti, V, Cr, Mn, Fe, Co, Ni; \emph{X}=Cl, Br, I) 
employing the parameter-free cRPA scheme. We found that in most of the metallic TM halides (in the non-magnetic state)
the Hubbard $U$ parameters for $t_{2g}$ or $e_{g}$ electrons are larger than 
3.0 eV, and the band widths $W$ are less than 1.0 eV. As a consequence, we find that $U/W>1$. So, 
these materials can be classified as moderately to strongly correlated systems.
The correlation strength in TM halides is much larger than the corresponding values in elementary 
TMs and TM compounds. Furthermore, using the calculated $U$ and $J$ values we discuss the stability 
of the ferromagnetic ordering within the Stoner model. The obtained Coulomb interaction parameters 
are important  both for a basic understanding of the physics of TM halides and for use in model 
Hamiltonians applied to describe electronic, magnetic, and optical properties of these materials.

The ferromagnetic state of $MX_3$ compounds can undergo  thermal
and quantum fluctuations. The thermal fluctuations are relevant to all of them. This provides a nice
experimental handle to study the role of the underlying magnetic background in transport and other properties
by tuning the temperature across the Curie temperature. This is a unique opportunity not present in generic
two-dimensional materials. For the compounds with small magnetic moment, the quantum fluctuations are
 expected to play a significant role. The effective  theory of small fluctuations around a magnetically ordered state is known as non-linear sigma model~\cite{FradkinBook}.
They describe spin-1 bosons known as magnons. Such bosons can mediate forces between the electrons, pretty much the same way
gauge bosons (photons) mediate the Coulomb interactions. The part of the interaction mediated by spin-1 bosons
cannot be screened~\cite{TsvelikBook}. Therefore our $MX_3$ compounds having smaller magnetic moments are expected to display many unexpected
correlation phenomena, such as non-Fermi liquids. The present paradigm for study of correlation physics in 2D systems,
unlike high temperature superconductors or heavy fermion systems are not buried in the bulk, but live in a true 2D layer.
They can therefore enjoy most of the control and functionalization methods developed in the context of graphene~\cite{KatsnelsonBook}.

Furthermore, in analogy to graphene nanoribbons~\cite{BreyFertig} where armchair ribbons offer low-energy one-dimensional bands,
the nanoribbons of TM halides are expected to serve as a new platform for strongly correlated
one-dimensional bands. Such bands are spatially separated from the other bands, and hence the fascinating Luttinger physics can be
studied using local probes. Also gapping out the edge modes in such ribbons~\cite{Louie} in the present correlated systems
can be an interesting framework for electronic applications that require an energy gap, albeit with substantial Mott character.

\subsection*{Acknowledgements}
E.S. and I.M. gratefully acknowledge funding provided by the 
European Union (EFRE, Project No. ZS/2016/06/79307).

\end{document}